\documentclass[epsfile,prc,aps,nofootinbib,showpacs]{revtex4}
\UseRawInputEncoding      %  needed only with the new Mac  (  a cause de matrix ???
\usepackage{xcolor}
\usepackage{graphicx}
\usepackage{epsfig}
\usepackage{amsmath}
\usepackage{amssymb}
\usepackage{gnuplot-lua-tikz}
\usepackage[normalem]{ulem}
\colorlet{darkgreen}{green!50!black}

\newcommand{\edit}[1] {\textcolor{red}{{#1}}}

\begin{document}

\title{Pionization  experiments in light nuclei}

\author{Slawomir Wycech}\email{Slawomir.Wycech@ncbj.gov.pl}
\affiliation{National Centre for Nuclear Research, Warsaw, Poland}

\author{Guillaume Hupin}\email{guillaume.hupin@ijclab.in2p3.fr}
\affiliation{Universit\'e Paris-Saclay, CNRS/IN2P3, IJCLab, 91405, Orsay, France}

\author{Jaume Carbonell}\email{jaume.carbonell@ijclab.in2p3.fr}
\affiliation{Universit\'e Paris-Saclay, CNRS/IN2P3, IJCLab, 91405, Orsay, France}

\date{\today}    % It is always \today, today,
%\date{September 2026}

\begin{abstract}
The zero-energy antiproton absorption in light nuclei is considered. 
The ratios $R_{n/p}$ of antiproton-neutron versus antiproton-proton capture rates
in the lightest nuclei $^2$H, $^3$He and $^4$He are analyzed, both from the experimental results and from some selected theoretical models. 
These models are used as key input in determining the level shifts and widths of the lowest states. We establish the capture orbit in $^3$He to be states of angular momentum $L=2$.
Several  N-\={N} optical potentials are used to compute the $R_{n/p}$ values for several heavy nuclei 
and the results are found to be consistent with the data. 
However, the old chamber experiments studying pionization in Helium,  are not consistent with models. 
Only the Paris potential is likely to succeed provided baryonia states predicted by this model are better understood. 
\end{abstract}

\maketitle

\section{Introduction}

Antiprotonic  atoms  and very low-energy  scattering  of antiprotons off nuclei  offer  a number of advantages for nuclear and low-energy particle physics:

First, the  studies of  atomic level shifts  and widths  allow us to test hadron-nucleon
scattering amplitudes  below the threshold, as one or both  particles are bound.  In this way one could trace 
exotic  baryonia (quasi-bound states in the two-body systems). 
Second, atomic level shifts and detection of atomic decay  modes
 of these systems are well established methods  to study the  structure of nuclear surfaces.  
Third, the nuclei and in particular light nuclei allow to test   antiproton-neutron  interactions which complement the knowledge of isospin  
dependence  of  $\bar{N}N$ forces.

In this work  we  concentrate on  a specific form of antiprotonic studies  
consisting in  measurements of charged pion spectra created in nuclear annihilation  of antiprotons (''pionization''). The  purpose of old and ongoing experiments  has been the detection of neutron haloes in nuclei. 
A pioneering experiment  on  Carbon target   measured  the total charge $Q$ of all  $\pi$   mesons  which managed to leave the  nucleus \cite{agn60}.  The charge distribution  $P[Q]$ was  determined  and the  usefulness  of the method for nuclear studies was proven.
Subsequent annihilation experiments in a freon chamber \cite{bug73}   found evidence of neutron halos in heavier nuclei.  A   ratio of antiproton annihilation on  neutrons versus annihilation on  protons  in  Carbon nucleus  was found in  a hydrogen chamber  \cite{wad76}. 

Nowadays,  a modern version of this type of experiments incorporating an  electronic detection  is   prepared at CERN  in PUMA project \cite{obe20}. It will study pionization in antiprotonic atoms of  radioactive nuclei.

The early experiments were not analyzed in terms of nuclear structure. The necessary knowledge of  $\bar{p} N$ and $\pi N$  interactions and the properties of antiprotonic atoms did  not  exist at that time.   
There were, and to some extent there are still,  two specific  questions  :\par
{\it i)}  the ratios  of $\bar{p}n$ and $\bar{p}p$  annihilation rates $R_{n/p}$ \par
{\it ii)}  the  atomic states of antiproton from which nuclear  captures  happens \par

Only with this knowledge, the nuclear region where the antiproton capture happens may be determined, and the neutron halo be quantified. 
Below we present briefly the status of current knowledge both experimental as well as theoretical.  
While the experiments in middle and heavy nuclei are well understood, the captures in light nuclei present problems.   
We try to understand these, check  several theoretical models of $\overline{N}-N $ interactions and suggest some improvements.
Reliable $\bar{N}N$ potentials are needed and the forthcoming experiments should be helpful. 

Apart from the pionization experiments, the majority of low-energy antiproton experiments involve X-ray measurements which yield atomic level width and shifts due to nuclear interactions of  $\overline{p}$. These are related to subthreshold  $\overline{p}-N $  amplitudes.  
A global analysis of the data is successful on a phenomenological level \cite{FRI07} but an analysis in terms of a theoretical model \cite{FRI15}  shows that nuclear attraction of antiprotons is not understood.  
On the other hand, the strengths of antiproton absorption is better controlled and an average best fit value of $R_{n/p}$ turns out to be about 1. 
Concerning  question {\it i)}  the  $R_{n/p}$  ratio  the  low-energy anti-protonic  experiments give  $R_{n/p}= 0.47(.04) $ in $^{3}$He \cite{ben91} $R_{n/p}= 0.48(10) $ in $^{4}$He \cite{bal87} and $R_{n/p}= 0.63 $ in C  \cite{wad76}, $R_{n/p}= 0.81(3) $ in Deuterium  \cite{biz74} and $R_{n/p} $  $\simeq 1 $ in heavy nuclei \cite{lub98}.  Sizable differences happen in light nuclei and reflect different nucleon-antiproton angular momenta  and  center-of-mass energies at the annihilation moment.  Determination of the latter becomes our main task.

 Each experiment  offers  a  specific  method  to answer the relevant  questions   {\it i)}  and  {\it ii)}.  The capture orbitals may be inferred  from rates of antiproton reactions leading to specific final nuclear  states  compared to total capture rates,  this  method was used in radiochemical  experiments \cite{lub98}. However, in these  of experiments the $R_{n/p} $ was  considered as a parameter to be taken from other sources.    For pionization studies ,  reference \cite{pis23}  offers  determination  the capture states from the charge spectra of emitted  mesons.  The lightest nuclei  allow for a kind of mixed procedure which we present and develop in this work.  It  is based on the  analysis of final state  mesonic interactions with the residual nuclei.  It allows also to extract $R_{n/p}$  directly from experimental data .
 
 In the pionization measurements    performed in Carbon, Deuterium and Helium   the $R_{n/p} $   were extracted  directly from  final mesonic charges,  but the capture states  remained unknown.   Unfortunately,  these ratios  are subject to final state interactions, apparently small in light but non negligible in heavier nuclei.  To find these  one has to know the capture states and the two uncertainties {\it i)} ,{\it ii)} have to be discussed jointly.
  
The present  paper is organized as follows:

Section II  introduces the pionization results,  i.e. the $\pi$-meson charge distributions obtained in  $^3$He and $^4$He  nuclei. 
These results are compared to those of the elementary reactions $ \bar{p} p \rightarrow$ mesons and $\bar{p} n\rightarrow$ mesons. 
Differences of the elementary distributions and those obtained in Helium  are used to establish the rates of  two final state reactions of the mesons: absorptions and charge-exchanges.
Next, these rates are calculated for three expected  antiproton-nucleus angular momentum states : S, P  and D.  Subsequent  comparison of  the phenomenological and the calculated results is used to determine the relative weights  of these  states in the  nuclear capture process.
%A subsection discusses neutron halo in $ ^6$He and  offers  expectations for future  measurements.

Section III discusses the predictions of three different  $\bar{N}N$ interaction models and tests these against  the $R_{n/p}$   and level shift  measurements. 
Some technicalities related to the final state description are presented in Appendix.

%%%%%%%%%%%%%%%%%%%%%%%%%%%%%%%%%%%%%%%%%%%%%%%%%%%%%%%%%%
\section{Pionic annihilation of antiprotons in  Helium}

We will first consider the measurement in  $^3$He \cite{ben91}, which offers the most precise and extensive information.
Pionic annihilation experiments in Helium were performed in a steam chamber \cite{bal87,ben91}.  
The results are compared here with elementary Hydrogen and Deuterium data from references \cite{kal80} and \cite{rie89}.
 
The primary reaction of antiproton-nucleon annihilation into $\pi$ mesons happens at the nuclear surface. The reactions on protons are known with  a considerable precision and are the basis of our analysis. 
The normalized distribution $\Pi(n)$ of the number of pions emitted  in $\bar{p}p$ annihilation 
 \begin{equation}\label{l1}
      \bar{p}  p \rightarrow  \pi_1 ...... \pi_n
\end{equation}
 (''pion spectrum'') covers charge and neutral  mesons  and extends from multiplicity $n=2$ till $n=8$ \cite{kle05}.  
Their value depends on the $\bar{p}p$ relative energy. The results corresponding to antiproton annihilation at rest 
presented in Table \ref{tabmultip}. They follow from chamber experiments in Ref. \cite{GHE} and are normalized to 100\% inside the error bars.

\begin{table}[ht]
\caption{ The pion multiplicities $\Pi(n) $ for  $\bar{p}p$  annihilations at rest [in $\% $].  }
\bigskip
\begin{tabular}{|    l   |    c    |    c    |    c    |    c     |    c       |      c     |      c        |}
\hline
 $n$                             & $2 $              & $3$             &     $4$            &    $5$               &   $6  $               & $7$              &    $ 8 $                                            \\ \hline
 $\Pi(n)$   & $0.38(0.03)$  & $7.8(0.4)$  & $17.5(3.0)$    &   $45.8(3.0)$    &  $22.1(0.15)$   &  $6.1(1.0)$   & $0.3(0.1)$                                   \\ \hline
\end{tabular} \label{tabmultip}
\end{table} 

\bigskip
The  measurement performed  in $^3$He  detects charged mesons and final charged  {\bf baryons}. The results of the reaction
\begin{equation}\label{l2}
\bar{p}  +     ^3He  \rightarrow  \pi_1 ...... \pi_n  +  N_b  \qquad n=0, \ldots, 7 
\end{equation}
consists of the frequencies of  charged pions and   the number $ N_b $ of charged baryons (predominantly protons, but these may  also be deuterons).  
To describe the frequencies we use symbol $ P_{N_b}[N_+,N_-]  $   where $N_{\pm}$  denotes  
respectively the number of positively  and negatively charged pions. 
Numbers  of neutral mesons are not determined. Each baryonic  sector  with  $N_b=0,1,2 $  is  presented below in  a separate table.

\begin{table}[ht]
\caption{Experimental  multiplicity distributions $P_1[N_+,N_-]$ in $^3$He  for  $N_b= 1$. Third column shows   $P_1$  as percentage of total events, \cite{ben91}.  Here,  $\alpha$ 
specifies  the number $\pi^+ \pi^-$  pairs given in the first column.  The four last columns give distributions  normalized to unity: in Helium, a weighted average of two  $\bar{p}p$ experiments (bold) and two  Hydrogen  measurements separately.}
\label{Nh1}
\bigskip
\begin{tabular}{| l |l|l|l|l |l|l| }
\hline
$N_+$ & $N_-$ & $P_1[N_+,N_-]$ &   $^3$He      & $ p\overline{p } $& ref. \cite{rie89}  & ref. \cite{kal80}  \\
\hline
0       & 0       &  4.31(.47)    &   0.0570 (.006)  &\textbf{0.0407(4) }  &  0.029       & 0.064(.004)   \\\hline
1       & 1       &  34.58(1.80)  &   0.4469(.0232)  &\textbf{0.4147(15)} &  0.398(.018)  & 0.423(010)   \\\hline
2       & 2       &  35.31(2.17)  &   0.4563(.0280)  &\textbf{0.4883(15) }& 0.527(.017)   & 0.469(.011)   \\\hline
3       & 3       &  3.16(.70)    &   0.00409(.009)  &\textbf{0.0447(5)} &  0.046(.01)    & .0441 (.003)  \\\hline
\end{tabular}
\end{table}

In  first approximation, $^3$He results given in Table \ref{Nh1} are due to initial ($\bar{p}p$) annihilations. 
Inspection  of the three last columns indicates that the differences of the two basic ($\bar{p}p$)  experiments are sizable, but still this sector offers the largest statistics  and the best known elementary reactions.  In further analysis we use the weighted average of the two  ($\bar{p}p$)  measurements.

The next $N_b$=0  sector is due  to  loss  of the final state proton  generated by  $\pi^-$  meson  produced in the dominant $N_b$=1 sector
\begin{equation}\label{cexm1}
       p ~ \pi^-  \rightarrow n ~\pi^0.
             \end{equation}
 Another process that links  these two sectors is an annihilation reaction,
\begin{equation} \label{abs}
      p~n ~ \pi^-  \rightarrow n ~ n , 
             \end{equation}
 but  we show in Appendix that it is negligible in the $^3$He case.

\begin{table}[ht]
\caption{ Experimental  multiplicity distributions  $P_0[N_+,N_-]$ in  $^3$He    $N_b= 0$ sector   (no final charged barions). }
\label{Nh0}
\bigskip
\begin{tabular}{| c |  c  |  c |} \hline
 $N_+$   & $N_-$   & $P_0 [N_+,N_-]$    \\  \hline
 1       & 0       &  0.59(.22)                      \\
 2       & 1       &  1.95(.46)           \\
 3       & 2       &  1.11(.32)           \\
 4       & 3       &  0.09(.09)           \\\hline
\end{tabular}
\end{table}

%%%%%%%%%%%%%%%%%%%%%%%%%%%%%%%%%%%%%%%%%%%%%%%%%%%%%%%
\subsection{Extraction of capture states  in  Helium }

The  purpose of this section is to establish the states from which  antiprotons are captured by the nucleus.  
To achieve this, we compare the charge spectrum obtained in Helium with the basic charge spectra  obtained in $\bar{p}p$  reactions. The differences are  attributed to final state  interactions of mesons and  residual nuclei. In a next step, a comparison of experimental and calculated differences is used to extract  the angular momentum of capture states.

The X-ray cascade in antiprotonic atoms indicates nuclear captures to happen at distant nuclear surfaces. Mesons are created in an initial $ \bar{p} N$ annihilations at small 0.8 $fm$ region around some point $\textbf{r}$. 
It is located inside a shell with a thickness of 2-3 fm and an average radius of about $R_{1/2}$ +0.6  fm, where $R_{1/2}$ is a half nuclear density radius.   
Next, these mesons propagate in all directions. Some cross the residual nucleus and interact. The processes of interest are mesonic absorptions and charge exchanges. These change the initial spectrum of pions  that would be emitted in a free  annihilation.   Comparison of two spectral distributions the  free one and one obtained in nuclei permits to establish the nuclear region where $\bar{p} $  was absorbed.   Detailed analysis is somewhat complicated by the initial mesonic spectra,  which contain also mesonic resonances: $\rho$ and $\omega $. However, the latter decay, within a one $fm $ distance,   into final $\pi$ mesons which are detected.  An analysis of reference \cite{wyc96} shows that angular averaging over mesonic trajectories allows us, to a good approximation, to discuss the $\pi$ mesons as born at an initial point  $\textbf{r}$.  
In the next section, the experimental spectra are discussed. Changes due to interactions with a final system are calculated in Appendix \ref{AA}. 

%%%%%%%
\subsubsection{Single charge-exchange }\label{firstorder}

The probability  of charge-exchange  reaction (\ref{cexm1}) per single meson is denoted by $\lambda$.  
The relation of   multiplicity probabilities  in sectors $N_b=0$ and $N_b=1$  is driven by $\lambda$  and one has for each number of $\pi ^+ \pi^-$ pairs.  :
\begin{equation} \label{M01}
P_0[N_+,N_--1]~ =~ \lambda \cdot~ N_-\cdot~P_1[N_+,N_-]
             \end{equation}
where $N_-$ is the   number of negative mesons in a given  initial channel in the $N_b$=1 sector.  
Data from Tables \ref{Nh1} and \ref{Nh0} allow to solve for  $\lambda$.  Corresponding values are given in the third column of Table \ref{M0to1}.

\begin{table}[ht]
\caption{Probabilities of transitions from the initial $N_b$=1 sector to $N_b$=0  sector.  
$K (K')$ are respectively the numbers of  $\pi ^+\pi^- $  pairs in the initial(final) sectors.  
$N_{\pi} $ are the average numbers of mesons involved in the  transition.   
The last three columns $\lambda^{(i)}_L$ shows results of  eikonal calculations performed for   three expected  angular momenta  of  the capture states. All $\lambda$ values on this table are multiplied by $10^2$.}\label{M0to1}
\bigskip
\begin{tabular}{|ll|c|c|l l l | }
\hline
   &transition   &  $N_{\pi} $ &$\lambda^{(i)}_{Exp}$& $\lambda^{(i)}_0$  &$\lambda^{(i)}_1$& $\lambda^{(i)}_2 $   \\ \hline
i=1 : &K=1$\to$K'=0 &   4         &  1.7(0.6)           &  4.47              & ~2.65           & ~1.40                \\\hline
i=2 : &K=2$\to$K'=1 &   5         &  2.7(0.6)           & 8.62               & ~5,19           & ~2.76                \\\hline
i=3 : &K=3$\to$K'=2 &   7         &  11.6(3.2)          &  13.6              & ~8.37           & ~4.52                 \\\hline
\end{tabular}
\end{table}

The  calculated $\lambda_L^{(i)}$  values depend strongly on the number of mesons involved in a given  transition.  As explained and calculated in Appendix, the charge-exchange reactions are dominated  by the $N \pi$  coupling to $\Delta(1233)$ resonance. It is the energy overlap of the $\Delta(1233)$  spectrum and  mesonic energy spectra  that determines $ \lambda$.  The mesonic energy spectra  depend on the total number of charged and neutral mesons emitted in a given channel.   Experiment \cite{ben91} detects charged mesons and the spectrum is given in Table \ref{MPT}. These serve to calculate   an energy spectrum  of  a single  meson.   
The  numbers of neutral mesons have to be imported from another experiment \cite{zem81}. The latter are shown  in Table \ref{MP}.  
 The spectrum in the  $MP[0,m]$  sector involving  small fraction of events  where no charge pions are emitted is not known. In calculations we use 
 $MP[1,m]$ instead. The related  uncertainty is minute in comparison to  uncertainties  of the  dominant sectors.
 
\begin{table}[ht]
\caption{ Probabilities of multi-meson state formation  in antiproton-proton annihilations  $MP[K]$ are obtained at rest in ref. \cite{rie89}.  The values related to annihilation on neutrons  $MN[K=0]$  are extracted from measurements taken at  1 GeV in ref.  \cite{zem81}. $K$ is the number of $ (\pi^+ \pi^-) $ pairs. }\label{MPT}
\bigskip
\begin{tabular}{|l|l|l|l|l|} \hline
K& 0          &  1              & 2                &   3   \\\hline
MP[K]& 0.029(0.005)  &  0.398(0.018) &  0.527(0.017)     & 0.046(0.010) \\ \hline
MN[K] & 0.055         &  0.734         &  0.189       & 0.022      \\ \hline
 \end{tabular}
\end{table}

\begin{table}[ht]
\caption{ Probabilities of multi-meson state formation  in antiproton-proton annihilations $MP[K,m]$, $K$ is the number of $ (\pi^+ \pi^-) $ pairs, $m$ is the number of $ \pi^0$ mesons. Results with error bars are proportional to experimental cross sections; results with no bars follow extrapolations made in ref. \cite{zem81}. Numbers in the $MP[K=0,m]$ column are not known.} \label{MP}
\bigskip
%\begin{tabular}{|l|l|l|l|r|}
\begin{tabular}{|l|l|l|l|l|} \hline
            &$MP[0,m]$ &  $MP[1,m]$     & $MP[2,m]$     &   $MP[3,m]$  \\ \hline
$m = 0 $   &        &  0.061(3)      &  0.144(7)     & 0.285      \\\hline
 $m=1   $  &        &  0.150(17)     &  0.522(53)    & 0.628    \\\hline
 $m=2   $  &       &  0.349 (16)    &  0.1807       & 0.083     \\\hline
 $m =3  $  &       &  0.333         &  0.1325       &  0        \\\hline
$m =4  $  &      &  0.073         &  0            &   0      \\\hline
$m = 5 $  &        &  0.034         &  &   0       \\ \hline
 \end{tabular}
\end{table}
The last three columns in Table \ref{M0to1} give the calculated values of $\lambda_L^{(i)}$,
following the method discussed in Appendix A.   The first  step of this calculation  finds  likely   densities  of antiproton - nucleon annihilation region  which is determined by overlap of the   atomic wave function and  a wave function of the  struck nucleon. These 
densities depend strongly on the angular momentum of the atomic state $L$. The next step  is a calculation 
of the chance that a meson  formed in a state of given $L$  changes  sign in the final state interactions. We follow  a prescription given in ref. \cite{wyc96} which shows that the  large energy released in the annihilation allows a closure approximation.  It permits to describe  residual nucleus in terms of optical potential with strength  related to charge exchange cross sections.  Average meson energy is about 450 MeV  high enough  to present the mesonic wave functions in terms of an eikonal.  As the calculated  values $\lambda_L^{(i)}$ depend strongly on  $L$  this gives a chance  to extract  the angular momenta of antiproton capture states.

Experimental  errors  of $P_0[N_+,N_-]$ are  large  and  to increase the confidence level  we  consider also the  second order transitions involving  two charge-exchange reactions.

%%%%%%%%%%%%%%%%%%%%%%%%%%%%%%
\subsubsection{Double charge-exchange  }

Consider two parallel  charge-exchange reactions  confined  to   the $N_b=1$ sector,
\begin{equation}\label{double}
     p ~ \pi^-  \rightarrow n ~\pi^0   ~~~~~~n ~ \pi^+  \rightarrow p ~\pi^0.
\end{equation}
These are of second order in $\lambda $  and  may show up only in the most precise parts of the data.  The  initial  antiproton-proton annihilation sector allows for that. For  an initial  state  with  $K$  ($\pi ^+, \pi^-$)  pairs, the final state  consist of  $K-1$ pairs.  
Comparison of the  primary $ (p,\bar{p})$ distribution with that obtained in Helium  shows  a loss of  mesons  in some  $K$  states and  a   gain of mesons in other K states.  The  loss  of rates is given by 
\begin{equation}\label{M01)}
 \delta ~ P_1[K]~ =~ (\lambda \cdot~ K )^2 \cdot~P_1[K]
\end{equation}

The inspection of Table  \ref{Nh1}  indicates that the largest loss   happens in  $K=$2   states. The difference of Helium and Hydrogen  results   is   $0.4563-\textbf{0.4883}= - 0.0320,  $  it is recovered  as  a gain  in  the   $K=$1 state \edit{:}
 $0.4469  -  \textbf{0.4147} = 0.0322 $.   This,  and other two $K$ states  allow us to extract  the charge-exchange parameters presented in Table \ref{lambda12}.  
 In this Table, we compare the rates $\lambda_{exp}$ extracted from the experimental data  by using the two methods: single charge  charge-exchange  and double charge-exchange. 
The latter are fairly close
to  results obtained from  the first order transition between $ N_b=1 $ and  $ N_b=0 $ sectors. However,  large errors arise  as the procedure involves subtractions of two large uncertain experimental quantities. In $^4$He  atoms  discussed in next sections  the double charge-exchange is the only detectable interaction.  That 
destroys the precision.

\begin{table}[ht]
\caption{Charge-exchange  parameters $\lambda_{exp}$  (in $10^{-2}$ units) obtained  from  one-meson charge-exchange (transitions  $ N_b=1  \rightarrow N_b=0 $) and two-meson charge-exchange  (transitions  $ N_b=1  \rightarrow N_b=1)$. K is the number of $\pi^+\pi^-$ pairs.}
\label{lambda12}
\bigskip
\begin{tabular}{| l |l|l|l|}
\hline
                  &  $K=3 $      &    $ K=2 $    &   $K= 1  $    \\ \hline
one exchange      &  1.7(0.6)   &  2.7(0.6)      & 11.(3.2)      \\ \hline 
two exchanges     &  1.17(1.5)  &  4.05(4.00)    &  9.72 (18.0)  \\ \hline
\end{tabular}
\end{table}
\subsubsection{Extraction of capture states}
To  calculate the  probabilities  of capture  from a given $L$ state - $w_L$ -    we use the  calculated  $\lambda_L^{(i)}$ values from Table \ref{M0to1} and 
assume a distribution saturated with a mixture of L=0, 1 and 2 states  
\begin{equation}\label{fitHe3)}
\lambda^{(i)}_{calc} =   w_{L=0} \cdot\lambda^{(i)}_{L=0}  +  w_{L=1}\cdot \lambda^{(i)}_{L=1} +  w_{L=2}\cdot \lambda^{(i)}_{L=2}. 
\end{equation}

Next we  determine their respective weights $w_{L }$ with a best fit to the three experimental transitions \ref{M0to1}. The detailed procedure of this fit is given in Appendix.
For $^3$He this yields  $w_{L=2}$=0.85+0.15-0.32, $w_{L=1}$=0.15+0.17-0.15 and $w_{L=0}$=0.00+0.10.
Comparison of calculated and phenomenological exchange parameters indicate  predominant captures from  $D$ states.
This result is consistent with the measurements by Schneider et al \cite{sch91} who  observe X-ray transitions to  $2P$ and $3D$ atomic states.
However, this measurement finds an  absolute yield of the lowest unperturbed $ 4\rightarrow 3 $ transition  to be only $35 \pm 9 \% $.  That shows importance of molecular effects which induce orbital mixing leading to  captures from  high $n$  states. For nuclear halo studies  this question is not relevant as in all cases the antiproton wave functions at nuclear distances take the form   $\Psi(r)  \sim r^L \cdot f(r) $  where $ f(r) $ describes nuclear effects  that are  roughly the same in all reactions.  The wave function normalization  is  irrelevant, unless  resonant mixtures of atomic states happen. 

Another problem  is:  how many events go by the in-flight captures ?   It is not known,  but again it is the $\Psi(r)  \sim r^L f $  behavior that matters,
The capture state distribution  obtained here is  close to calculations of Reifenroether et al \cite{reif88} who  find  $8\% $ from S-waves $49\% $ from P-waves and  $53\% $ from D-waves in gas at 1 atm pressure.  
The  X-ray  measurement  was done in a gas at  600 mb, while the pionization experiment   was done at normal pressure.  The results by PUMA will be different anyway as the  atomic capture will happen in vacuum. What matters is the detection of  charge baryons allows to  fix the  capture states.

The  Helium experiments offer  also some additional information.  Table \ref{Nh2} presents  the multiplicities following initial antiproton-neutron  annihilations.  However, there are two problems that hinder  our  analysis: larger experimental errors and lower capture rates. No significant gain of precision may be obtained.

\begin{table}[ht]
\caption{ Experimental  charge multiplicities $P_2$ in $^3$He  for  $N_b= 2$ ($\bar{p}n $) primary capture.  
Third column shows   $P_2$  as percentage of total events, \cite{ben91}. The four  last columns give normalized to unity  fractions: in  Helium,  a weighted average  fractions in $\bar{p}n$ (bold) and   of two  deuterium  measurements
performed  in  corresponding references. }
\label{Nh2}
\bigskip
\begin{tabular}{| l |l|r|r|r|  l  |  l | }
\hline
$N_+$ & $N_-)$ & $P_2[N_+,N_-]$  & $norm$  & $\overline{p}n $ & \;\; \cite{rie89}         &  \;\;\cite{kal80}  \\
\hline
    0  & 1       &  2.72(.50)     &   0.1440 (.026)    & \textbf{0.116}    & 0.065 & 0.168(.008)   \\\hline
1       & 2      &  12.13(1.22) &   0.6431(.06468)  & \textbf{0.667} & 0.733 & 0.595(.014)   \\\hline
2       & 3      &  3.86(.770)  &   0.2047 (.0408)  & \textbf{ 0.209}  & 0.189 & 0.230(.009)   \\\hline
3       & 4      &  0.15(0.11)   &  0.0079 (.0059)  & \textbf{0.041}    &  0.0121 & 0.07(.020)   \\\hline
\end{tabular}
\end{table}

%%%%%%%%%%%%%%%%%%%%%%%%%%%%%%%%%%%%%%%%%%%%%%%%%% 
\subsection{Captures in $^4$He}

%%%%%%%
\subsubsection{ Pionization in $^4$He }

The $^4$He pionization  experiment  does not detect  charged baryons, apparently due to the strong binding of the latter.     
The  former sectors $N_b=1$  and $N_b=2$ that is  initial  $(p\bar{p})$ captures  and 
 ($n\bar{p}$)  captures  are lumped together   and no equivalence to the  $N_b=0 $  sector  exists \cite{ben91}. 
 The charged pions distributions for $^4$He are simply denoted by $P$.
  Again, as  the  ($p\bar{p}$) data are  more precise  that the $(n\bar{p})$ data   and we consider captures of final zero charge that predominantly are due to   the initial ($p\bar{p}$)  annihilations. 
 As   the  analysis in terms of single charge-exchange reaction is not easy,  one has to use  the double exchange transitions (\ref{double})  and  primary  $p\bar{p}$  capture data 
 shown in Table \ref{He4a}.  The related charge-exchange  rates are collected in Table \ref{He_4b}.

\begin{table}[ht]
\caption{Experimental  multiplicity distributions $P$ in $^4$He \cite{ben91},   $K $  is the  number of  $\pi ^+\pi^-$  pairs. The  rightest column gives the  primary $(p\bar{p}) $ fractions.}\label{He4a}

\bigskip
%\begin{tabular}{| l | l | l | l | l | }\hline
\begin{tabular}{| l | l | l | l | l | }\hline
$K$ &  $N_+$ & $N_-$ & $P$  \;       & $P$ \; $\bar{p}p$  \\
\hline
0&  0       & 0       & 0.0496(.0052)    &    \textbf{0.0407(4) }     \\\hline
1&  1       & 1       & 0.4664(0.0165)  &    \textbf{0.4147(15)}    \\\hline
2&   2       & 2       &  0.4383(0.0252)&    \textbf{0.4883(15) }  \\\hline
3&  3       & 3       &   0.0420(0.0142)&  \textbf{0.0447(5)}   \\\hline
\end{tabular}
\end{table}

\begin{table}[ht]
\caption{
Probabilities of charge-exchange transitions $\pi^- p \rightarrow \pi^0 n $  and   $\pi^+ n \rightarrow \pi^0 p $  in $^4$He. 
$K (K') $  are  numbers of  ($\pi ^+ \pi^-$)  pairs in the initial (final) state.  
$N_{\pi} $ are  the average numbers of mesons involved in the  transition. The numbers given in column three  are experimental  $\lambda$  numbers. 
The last column gives $\lambda$ obtained with  eikonal calculations performed for  the   three expected  angular momenta  of  the capture states. 
All $\lambda $ values are multiplied by $10^2$}.\label{He_4b}

\bigskip
\begin{tabular}{|ll|c|l|lll|}
\hline
   & transition   &  $N_{\pi} $ & $\lambda^{(i)}_{Exp}$ & $\lambda_0^{(i)}$ & $\lambda_1^{(i)}$ & $\lambda_2^{(i)}$   \\ \hline
i=1& K=1$\to$K'=0 &   4         & 10.3(9.1)             & 11.4              &  8.45             & 5.7            \\\hline
i=2& K=2$\to$K'=1 &   5         & 11.3 (8.7)            &  18.4             & 13.9              & 9.5                \\\hline
i=3& K=3$\to$K'=2 &   7         &  5.8(14.0)            & 25.1              & 19.3              & 13.4          \\\hline
\end{tabular}
\end{table}
 
Table \ref{He_4b} indicates  a dominant $L$=0  captures  happening in  $^4$He  atoms which,  according to Appendix \ref{Fit}, gives the populations  
$w_{L=0}$=0.77+0.13-0.14,   $w_{L=1}$= 0.00+0.18 and  $w_{L=2}$=0.23+0.26-0.23.   Possibly, it is due to  stronger  mixing of  atomic states  generated by Stark effect  in the early stage  of cascade and absorptions from $S$ states and high main quantum number $n$.  The findings of Schneider  \cite{sch91} 
indicate  $65\%$ of absorptions  to  happen in high $n$ possibly $S$ states and  such result is not excluded. However, the difference of  $^3$He  and $^4$He  makes  it  hard to accept.  More likely it is related to uncertainties of  our method which requires  final state  effects  larger than experimental errors. 
An estimate better than that leading to Table \ref{He_4b} should  consider inverse reactions    
$  n ~\pi^0  \rightarrow    p ~ \pi^-   , ~ n ~ \pi^+  \rightarrow p ~\pi^0 $. 
This  approach would however involve much uncertain  data: the  rates of  $ \pi^0$ formation  and  $\bar{p}n$  pionization spectra, and hence  a reliable calculation cannot be fulfilled.    
The extraction of capture orbits will have to be  achieved  by other methods, possibly  via  comparison of  $^4$He and $^3$He  atomic cascade calculations. 

%%%%%%%%%%%%%%%%%%%%%%%%%%%%%%%%%%%%%%%%%%%%%%%%%%%%%%%%%%
%   COMMENT   S :  we use the same  S,PD, letters For atomic  and n-nbar momenta . I tried to put atomic inn italics
\section{Testing  $\bar{N}N$ interaction models}

%%%%%%%%%%%%%%%%%%%%%%%%%%%%%%%%%
\subsubsection{ Extraction  and understanding  the  $R_{n/p}$ value}

Table \ref{Nh2}  presents pionization in  $^3$He initiated by  antiproton-neutron absorption. Let us notice that differences between Helium and $\bar{p}n $ spectra are much smaller then experimental errors.  Hence, the type of analysis performed in the  last section would produce very uncertain results.  Still, these results are essential, showing the  isospin dependence of the annihilation. 
It yields  the ratio of  $\bar{p}n$  and  $\bar{p}p$ capture rates $R_{n/p} = 0.465(.0415)$, which is very important in the nuclear  structure studies. This  
value is obtained in ref. \cite{ben91} on the basis of experimental results given in  Tables \ref{Nh1} and \ref{Nh2}, 
 as a ratio of $N_{b=2}$ and ($N_{b=1}$ + $N_{b=0}$) events.  Below, we check it  against  corrections due to   $N_{b=1} \leftrightarrows N_{b=2} $  balance.
 The transition  $N_{b=2} \rightarrow N_{b=1}$  and the return one are given by reactions
\begin{equation} \label{cexm}
       p ~ \pi^-  \rightarrow n ~\pi^0; ~~ n ~\pi^0  \rightarrow p ~ \pi^- .
\end{equation}
The balance of the  two sectors   depends on  $P[N_+,N_-]$  given in Tables \ref{Nh1}, \ref{Nh2}  and  the content of neutral mesons given in table \ref{MP}. 
The values of experimental $\lambda $ obtained in section \ref{firstorder}  and formulas analogous to eq.( \ref{M01}) yield  a minute  and insignificant  enhancement $0.001(.05)$ which leads to   $R_{n/p} =0.467(0.035).$

An open problem  is  understanding of  the $R_{n/p}$ in terms of nucleon - antinucleon potentials. Below we analyze several results.   
The first question is:  what is the relevant energy in the $\bar{N}N$  system that leads to the experimental value of $R_{n/p}$  ?   To answer we follow reference \cite{loi20} which discusses light anti-protonic atoms as  quasi three body systems :
antiproton (1), struck nucleon (2)  and a  residual nucleus (3).  This allows to determine  energies in the $\bar{N}N$   center  of mass system  as 
\[  E_{CM}  = 2M_N - E_S - E_A - E_{REC} \] 
where: $E_S$  is  a  separation energy   of the nucleon,  $E_A$  is  a (small)  antiproton binding energy and $E_{REC}$ is a recoil energy of  $\bar{N}N$ pair  with respect to  the residual nucleus. 
Calculations performed in  ref. \cite{loi20} yield the  values  shown  in table \ref{Esub}.  
It turns out that  $R_{n/p}$  are generated  at subthreshold energies of about $ - 34 $ MeV in $^4$He   and  about $ - 14 $ MeV in $^3$He. 

The second question is: what are spin states of the $\bar{N}N$  system  at the annihilation ?.  In atomic D waves the  result is given by  expressing  the atomic  spin state  in terms of  internal spin of the $\bar{N}N$.   The absorption in this state is essentially given by Born term and we operate on spin averaged amplitudes. This allows  S,P,D internal angular momenta, but  D-waves are negligible   and  the weights of S and P waves is calculable  by  3- body kinematics  with  known  wave functions.  Ref. \cite{loi20} gives  the relative weights of the internal $\bar{N}N$  states  as  shown in table \ref{Esub}.

\begin{table}[ht]
\caption{ Separation and recoil energies in $L$= 2 atomic states.  The second column gives averages of proton and neutron separation energies  {MeV}. Third column gives average recoil energies [MeV], numbers in parentheses are approximate widths of recoil energy distributions. The last two  columns presents weights  $w(L)$ of $ N,\overline{p } $ captures in relative S and P  waves obtained with Paris potential.}\label{Esub}

\bigskip
\begin{tabular}{  |   c  |             c    |                   c          |         c        |         c                  |  }\hline
                                 & $E_S$     &  $<E_{REC}> $     &    w(S)     &    w(P)               \\ \hline
$ ^3$He                    &  6.6          &  7.3(1)                   &   0.38      &    0.62                                  \\\hline
$^4$He                     & 20.2         &  13.7(10)               &   0.41      &    0.59                 \\\hline
\end{tabular}
\end{table}

The last question is to find annihilation rates in  S and  P waves  at  subthreshold energies. 
According to ref. \cite{loi20} we define the off-shell below-threshold $\bar{N}N$ T-matrix  in a given partial wave $\alpha\equiv\{TLSJ\}$ as
\begin{equation}\label{fbth}
t_{\alpha}(E,r)= {1\over (2L+1)^2}\; {1\over 4\pi} \; { v_{\alpha}(r)  \left[  \hat{j}_L(qr) 
 +  \chi^+_{\alpha}(r)   \right]\over \hat{j}_L (r) }   
 \end{equation}
where
\[ v_{\alpha}(r) = \; {2\mu_{N\bar N}\over\hbar^2} V_{\alpha}(r) \]
is the the N\={N} reduced potential in the partial wave  $\alpha$, $\mu_{N\bar N}=m_N/2$ is the N\={N} reduced mass, 
$\hat{j}_L(z)$ is the reduced regular spherical Bessel function \cite{Abramowitz_1970}, 
$\chi^+_{\alpha}$ is the outgoing part of the corresponding 
reduced regular radial scattering solution $u_L\equiv\hat{j}_L(qr)+ \chi^+_{\alpha}(r)$ \cite{Taylor_1972}.

This definition is such that the integrated quantity
\begin{equation}\label{aLE}  
t_L(E)= 4\pi\int_0^{\infty} r^2dr \;  t_L(E,r) r^{2L} =  4\pi \int_0^{\infty} dr \; r^{2L+2}  \;  t_L(E,r) 
\end{equation}
coincides at $E$=0 with the usual scattering length: $t_L$(E=0) $\equiv - a_L$ \footnote{Using the standard nuclear convention $q^{2L+1}\cot\delta_L(q)= -{1\over a_L}+ o(q^2)$}. 
For $E<0$,  $t_L(E)$ constitutes a natural extrapolation  of the usual scattering amplitude to  subthreshold energies.

\begin{figure}[h!] 
\begin{center}
\includegraphics[width=8.5cm]{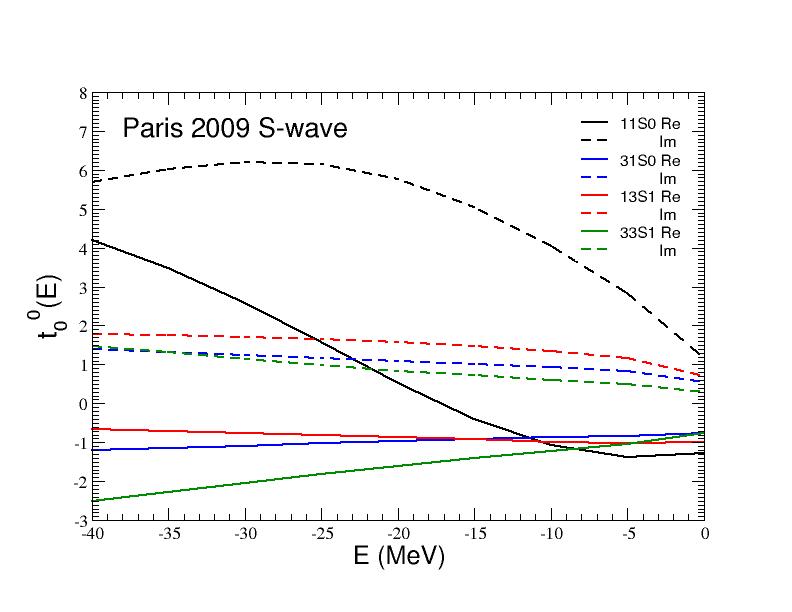}%
\includegraphics[width=8.5cm]{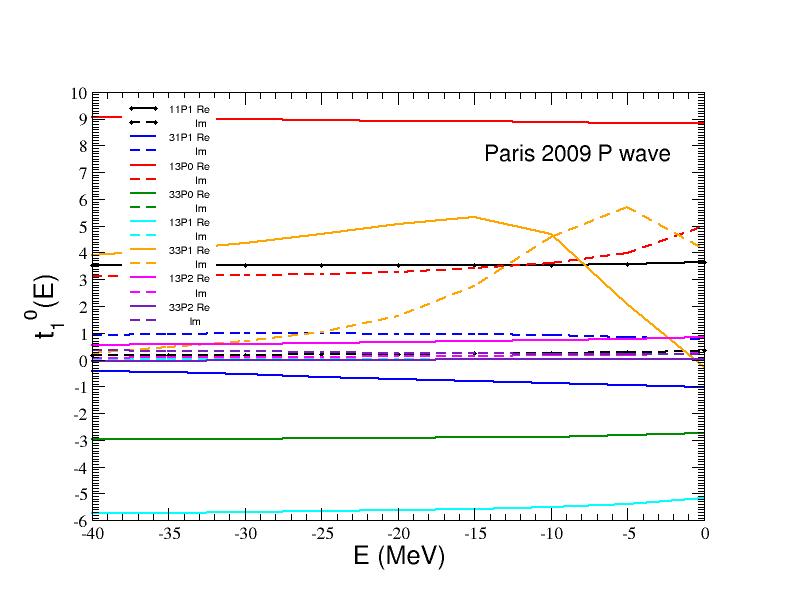}%
\end{center}
\vspace{-0.7cm}
\caption{Below threshold amplitudes for the Paris potential \cite{PARIS_PRC79_2009}.}\label{Fig_A0_A1_Paris}
\end{figure}

\begin{figure}[h!] 
\begin{center}
\includegraphics[width=8.5cm]{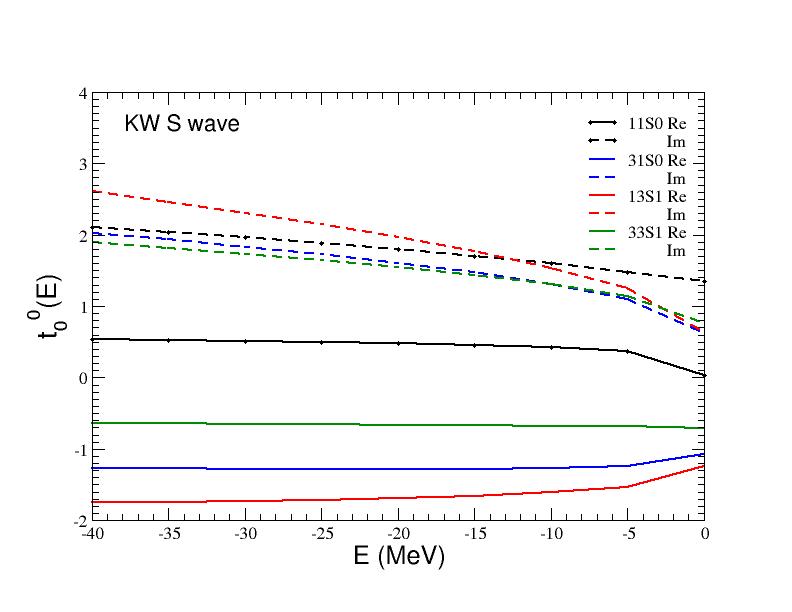}%
\includegraphics[width=8.5cm]{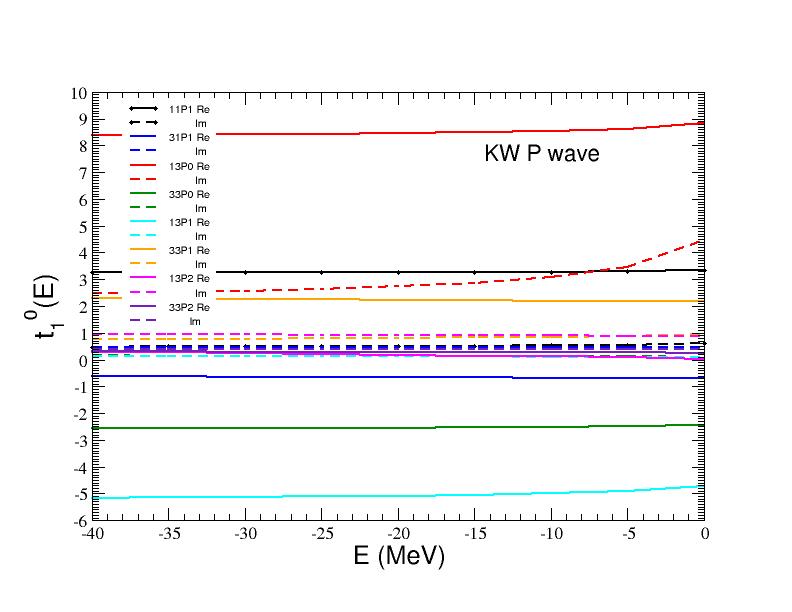}%
\end{center}
\vspace{-0.7cm}
\caption{Below threshold amplitudes for the Kohno-Weise potential \cite{KW_NPA454_1985}.}\label{Fig_A0_A1_KW}
\end{figure}

\begin{figure}[h!] 
\begin{center}
\includegraphics[width=8.5cm]{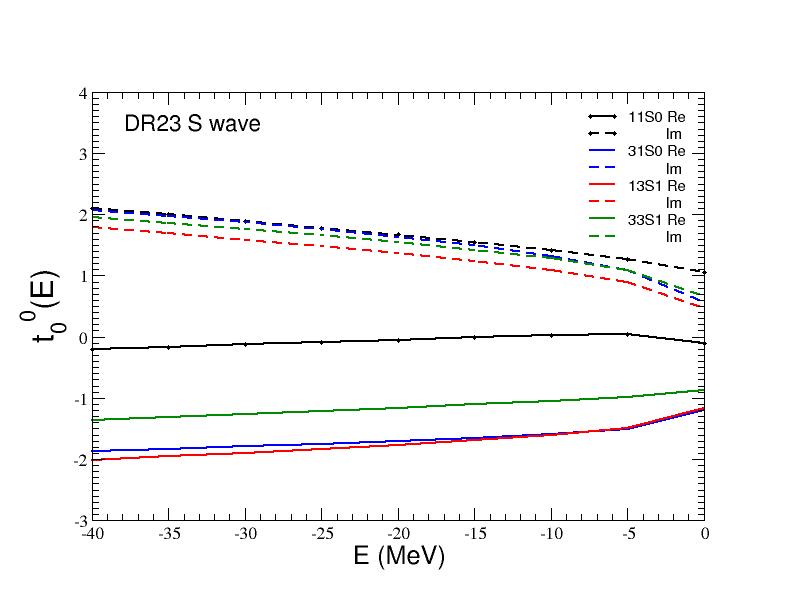}%
\includegraphics[width=8.5cm]{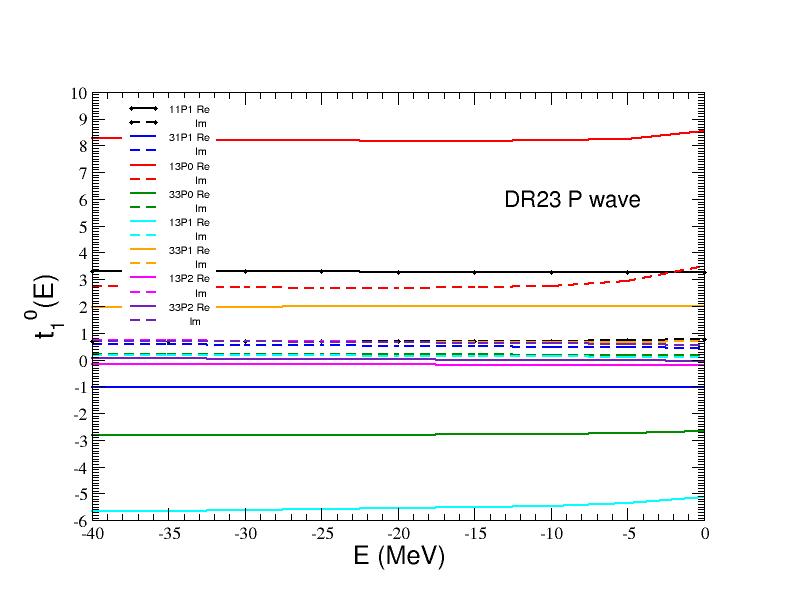}%
\end{center}
\vspace{-.7cm}
\caption{Below threshold amplitudes for the Dover-Richard model \cite{DR2_RS_PLB110_1982}.}\label{Fig_A0_A1_DR23}
\end{figure}

It is also interesting to introduce the "moments of the off-shell T-matrix" by
\[  t^{(2K)}_L(E)   \equiv  4\pi \int_0^{\infty}  dr  \; r^{2L+2} \;t_L(r,E) \; r^{2K} \]
where for K=0 one has $t^0_L(E)=t_0(E)$.
We have computed the S- and P-waves amplitudes  $t^0_L(E)$ for 
the  $\bar{N}N$  optical potentials in configuration space that we have examined in \cite{Comparison_EPJA59_2023},
i.e.  Paris 2009 \cite{PARIS_PRC79_2009}  	
Dover-Richard model 2 \cite{DR2_RS_PLB110_1982} 	
and Kohno-Weise model \cite{KW_NPA454_1985}.
The results for the L=0 ($t^0_0$) and L=1 ($t_1^0$) 
amplitudes  as  functions of the c.o.m. energy are displayed in Fig. \ref{Fig_A0_A1_Paris}
for Paris potential, in Fig. \ref{Fig_A0_A1_KW} for Kohno-Weise (KW)
and in  Fig. \ref{Fig_A0_A1_DR23} for Dover-Richard model 2 (DR23).
Different colors  correspond to different partial waves in the isospin basis.
The real parts  are in solid lines and the imaginary pars in dashed lines.
Notice that for all partial waves, and in all the energy range, one has Im$[t_L(E)]>0$, which corresponds at E=0 to Im$(a_L)<0$.

\vspace{-0.cm}
\vspace{-0.cm}
\vspace{-.cm}

For the S-wave (left panel) the singular character of the Paris potential
is manifested, due to the presence of a near-threshold quasi-nuclear state in the $^{11}$S$_0$ channel.
This state is absent in KW and DR model and hence the amplitudes have a smooth behavior
as a function of the energy, in sharp contrast with Paris.

For P-waves (right panel) a strong energy dependence in Paris potential is displayed by the $^{33}$P$_1$ 
partial wave, also due to the narrow bound state. A common feature of the three models is the large
values of the $^{13}$P$_0$ amplitudes.

\begin{figure}[h!] 
\begin{center}
\includegraphics[width=6.2cm]{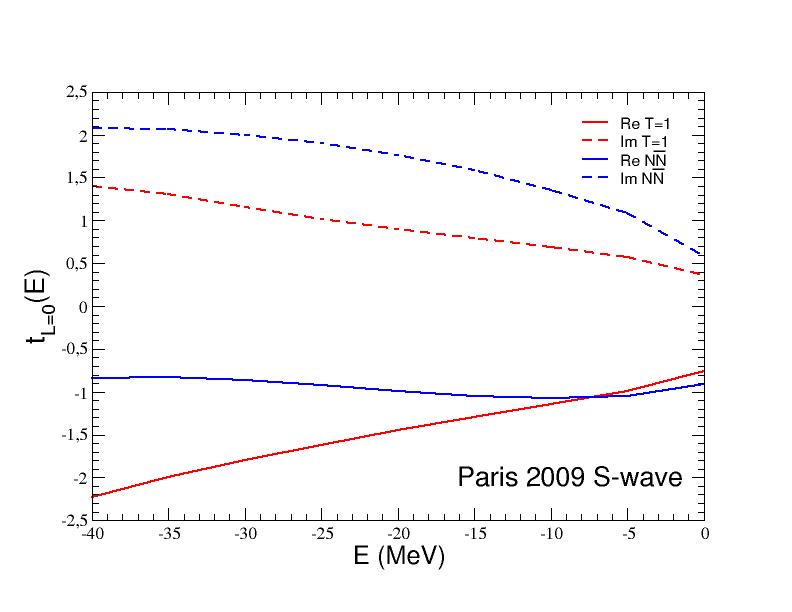}%
\includegraphics[width=6.2cm]{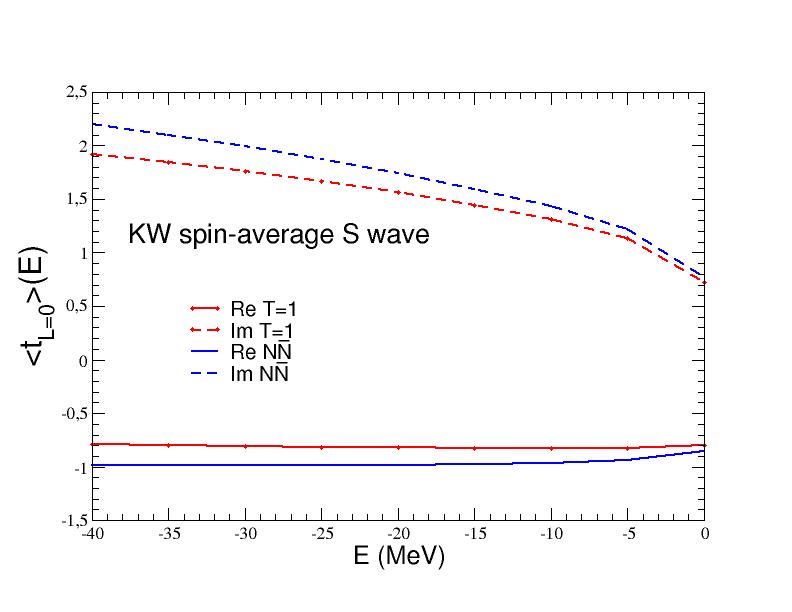}%
\includegraphics[width=6.2cm]{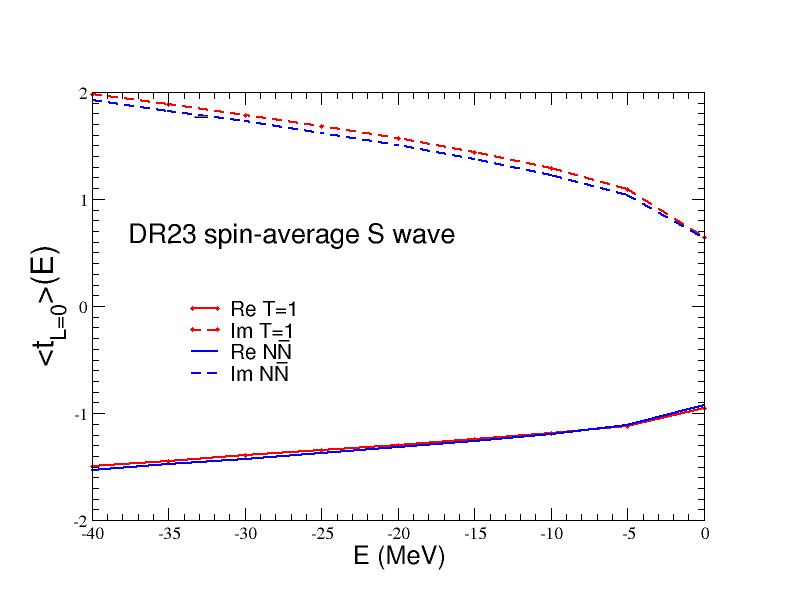}%

\vspace{-0.7cm}
\includegraphics[width=6.2cm]{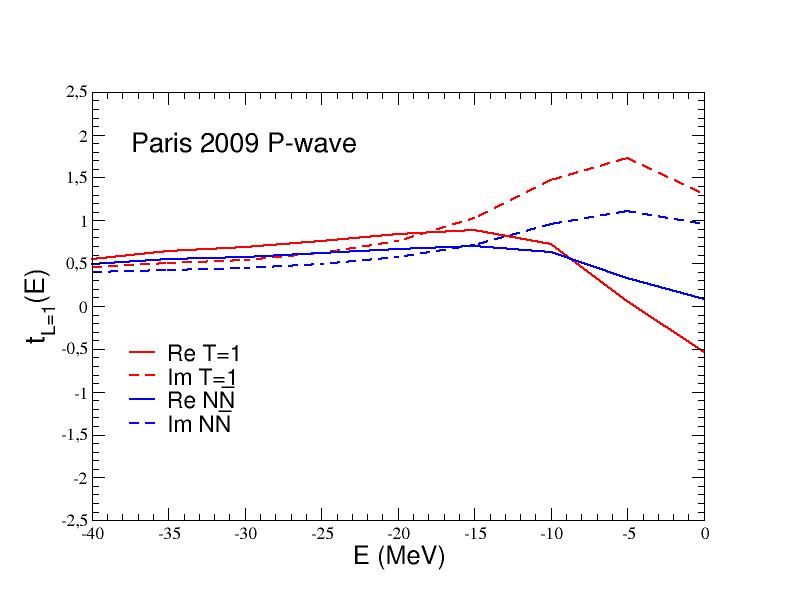}%
\includegraphics[width=6.2cm]{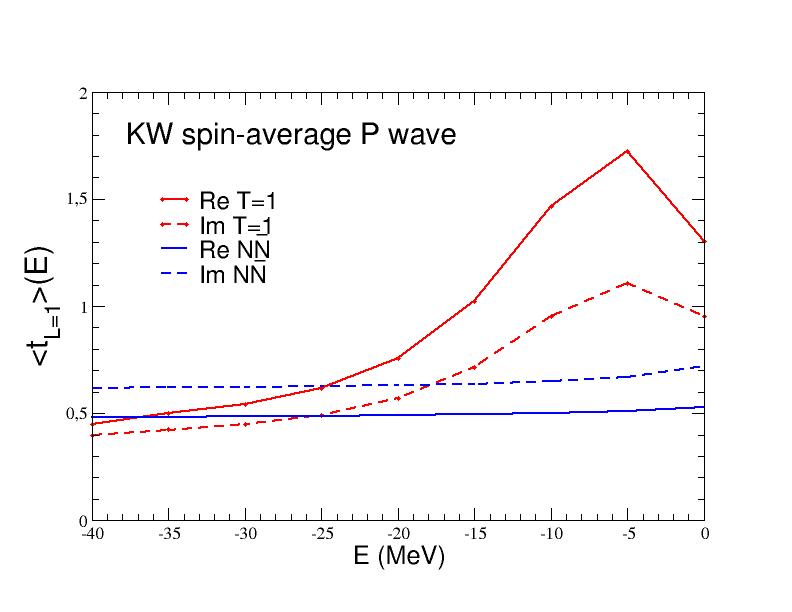}%
\includegraphics[width=6.2cm]{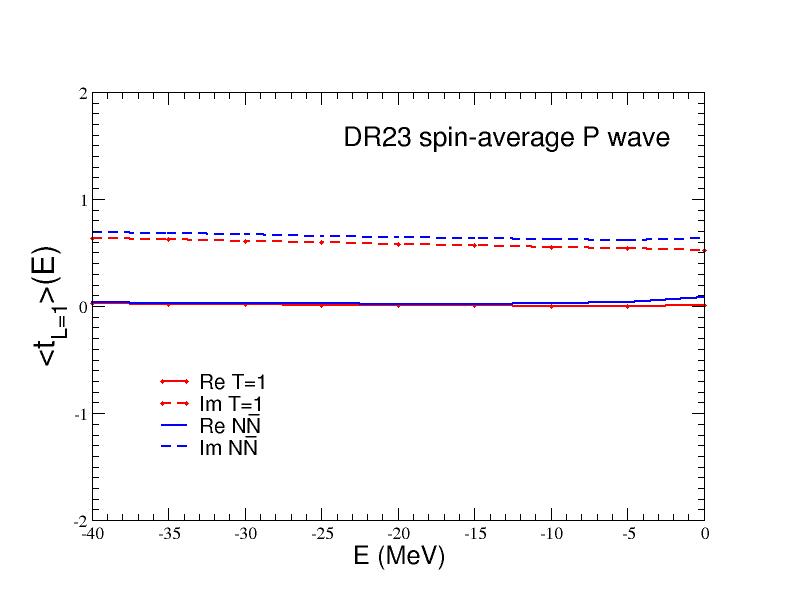}%
\end{center}
\vspace{-0.7cm}
\caption{Spin-averaged values of S- (upper panel) and P-waves (lower panel)  
below threshold amplitudes for the T=1 and $\bar{N}N$ states computed with  Paris, KW and  Dover-Richard \cite{DR2_RS_PLB110_1982} optical models.}\label{Fig_Spin_Av_SP}
\end{figure}

Our aim is to compute -- for the lowest angular momentum states -- 
the ratio $R_L(E)$ between the imaginary part of the amplitudes (\ref{aLE})  for T=1  [$t_L(T=1)$] and the isospin-averaged values $t_L(\bar NN)$ defined as
\begin{equation}
    t_L(\bar NN)= \frac{1}{2} \Bigl[ t_L(T=0)+t_L(T=1) \Bigr] 
\end{equation} 
Still, for an easier comparison with the available experimental data it is interesting
to take the "spin-averaged " values of the amplitudes, i.e.
\begin{eqnarray}
<t_0>  &=&  \frac{1}{4}   \Bigl[   t_0(^1S_0)+ 3\; t_0(^3S_1)                        \Bigr]              \\
<t_1>  &=&  \frac{1}{12}  \Bigl[   t_1(^3P_0)+ 3\; t_1(^1P_1)+ 3\; t_1(^3P_1) +  5\; t_1(^3P_2) \Bigr]    
\end{eqnarray} 
for S waves and P-waves respectively. Corresponding results are displayed in Figs \ref{Fig_Spin_Av_SP}.

%\vspace{-0.7cm}

\begin{figure}[h!] 
\begin{center}
\includegraphics[width=8.5cm]{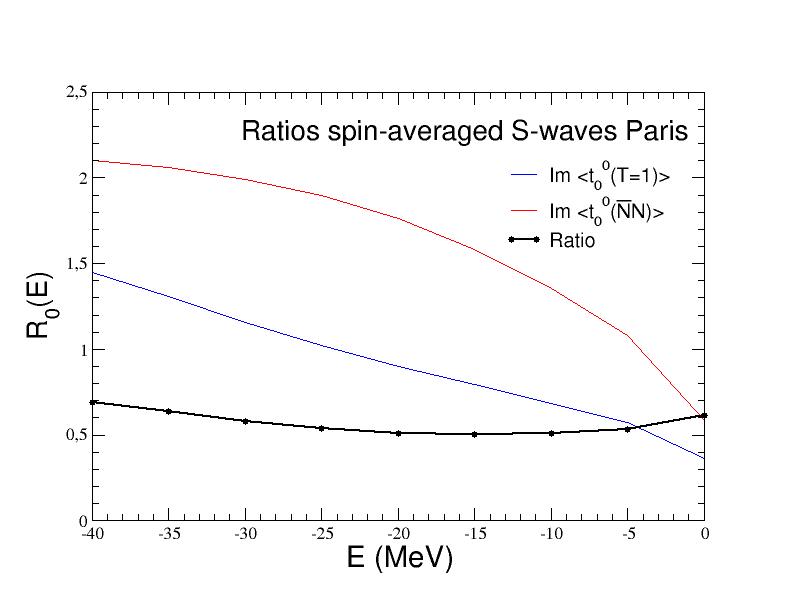} 
\includegraphics[width=8.5cm]{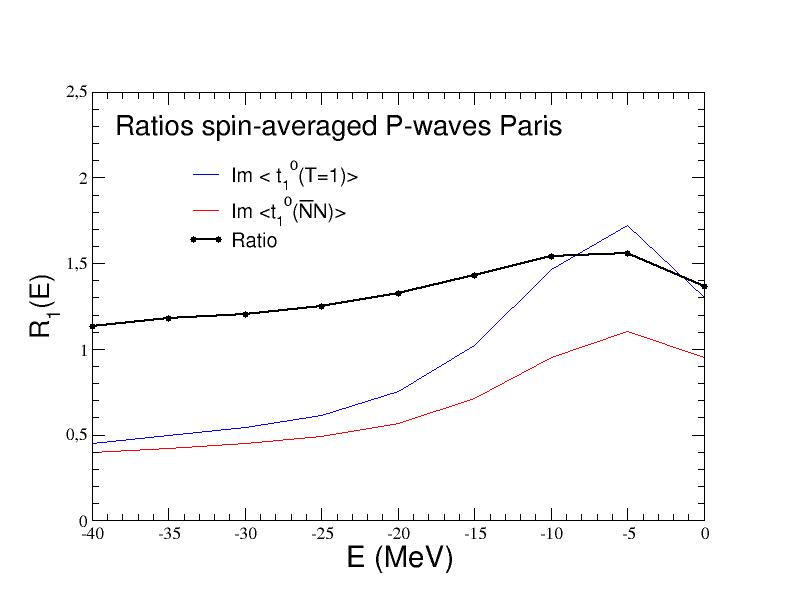} 

\vspace{-.8cm}
\includegraphics[width=8.5cm]{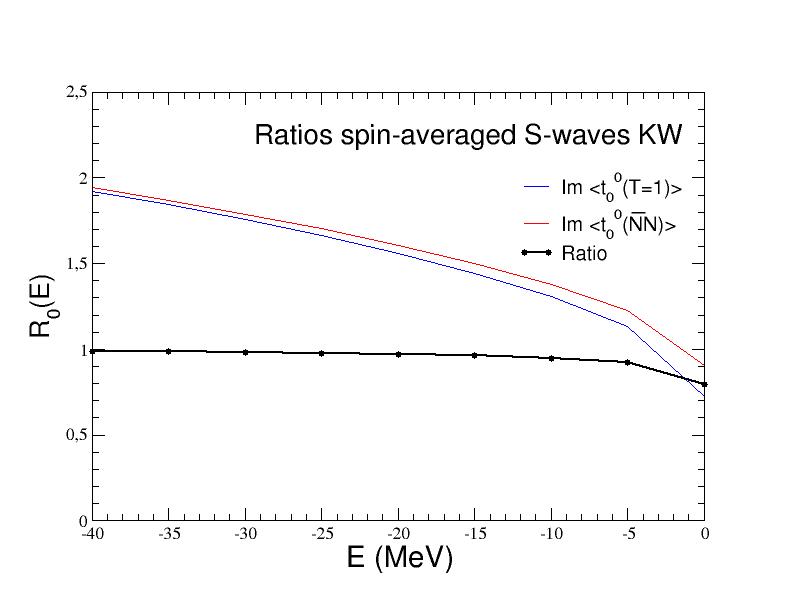} 
\includegraphics[width=8.5cm]{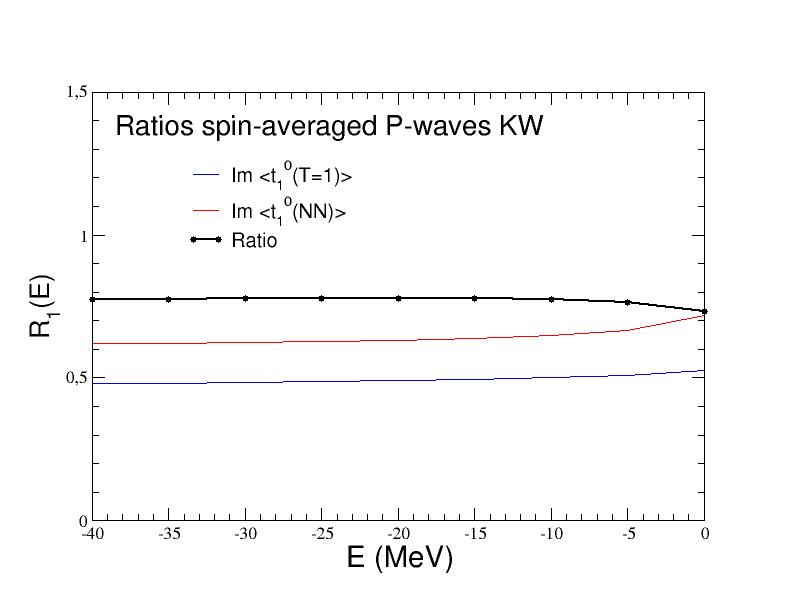} 

\vspace{-.8cm}
\includegraphics[width=8.5cm]{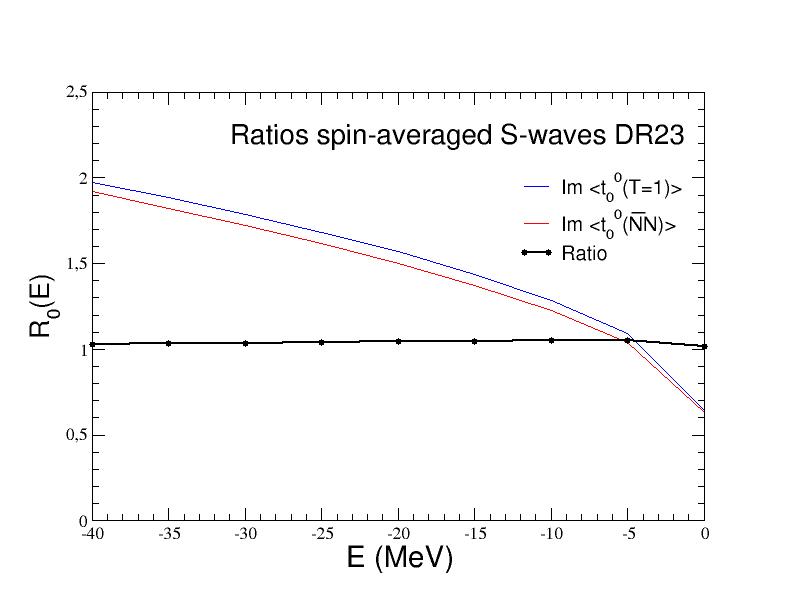} 
\includegraphics[width=8.5cm]{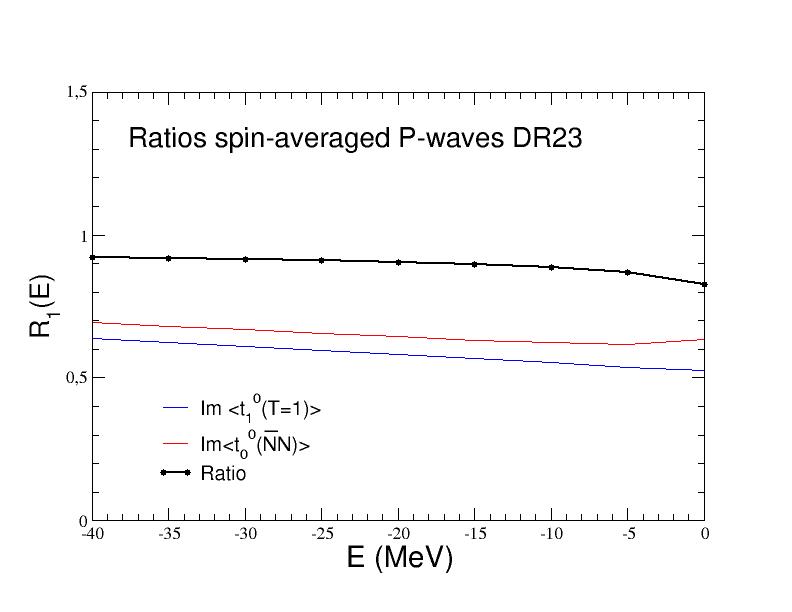} 
\end{center}
\caption{Ratios (\ref{Ratio}) for S-  ($R_0$) and P-waves ($R_1$) as  functions  of the center of mass energy, obtained
with Paris (upper panel), KW (medium panel) and DR23 (lower panel) potential}\label{R_Paris_KW_DR23}
\end{figure}

One is then left to compute
\begin{equation}\label{Ratio}
R_L(E)= {  {\rm Im} \left[ <t_L^0 ( {\rm T=1} )> \right] \over {\rm Im} \left[ < t_L^0 (  \bar{N}N    ) >\right]     } 
\end{equation}
The results of (\ref{Ratio}) for the Paris and KW potential are displayed in Fig. \ref{R_Paris_KW_DR23}.
Results presented in Fig. 4, are summarized in table \ref{rnp}. The elementary ratios for  Sand P-waves follow the spin-averaged subthreshold absorptive amplitudes calculated for three models.  
Next these were weighted by probabilities of  S- and P-wave captures, W(S) and W(P), given in table \ref{Esub} .  D  wave  contributions are negligible.

\begin{table}[ht]
\caption{Ratios of average   $ \bar{p}n $  and  $\bar{p}p$ captures in $^4$He  calculated with several  $ N\bar{N } $ potentials.   The  S, P,D  are atomic angular momenta . The last column  gives  the weighted average based on calculated  
contributions of  nucleon-antinucleon  states given in table  \ref{Esub}                }. 
\label{rnp}
\begin{tabular}{|l|l|l|l|l| }
\hline
$ ^3$He       & $R(S)  $        &    $ R(P) $  &   $ R(D) $ &  $ R_{n/p} $   \\ \hline
Paris         &  0.51           &     1.40    &    1.00         & $ 0.72$       \\\hline
K-W           &  0.99           &     0.80    &     0.77      &  $ 0.89 $       \\ \hline
 D-R          &  1.04           &     0.90   &     $ 1.17$ &  $ 0.95$         \\ \hline     
 $ ^4$He      &                 &             &             &                 \\  \hline
Paris         &  0.62          &     1.18     &    0.72         & $ 0.94 $           \\ \hline
K-W           &  1.00           &     0.88     &    0.78         & $ 0.89 $       \\\hline
 D-R          &  1.03           &     0.91   &   1.18       &   $ 0.99$         \\\hline
\end{tabular}
\end{table}

With the captures from atomic $D$-states, it is the P-wave $\bar{N}N$ absorption which dominates the $R_{n/p}$  ratio.   
Inspection of Figure 4   and  Table XIV 
shows inconsistency of all models with the experimental $ ^3$He   result of $0.47(.04)$. 
Also, the experimental $ ^4$He   result of $0.48(.10)$  is not reproduced by any potential.  On the other hand, all models are consistent with 
the pionization and X-ray experiments in heavier nuclei \cite{lub98}, where $R_{n/p}$ are determined predominantly by $P$-wave interactions.  

At present, Paris model is the only potential that offers small $R_{n/p}$ values, and only in the S-wave. It comes  
as a result of an I=0 baryonium state formed in the $N \bar{N}$ system. 
As shown in Fig. 5, the presence of this quasi-bound state pulls the ratio down to the values obtained in Helium. The S-wave baryonium in question has an independent confirmation in $J/\psi$ decays \cite{DED4}. It also offers an explanation of X(1835) state \cite{DED4}.  On the other hand, the 
$R_P$  ratio in P-wave spoils consistency with the Helium results. It is due to a  P-wave baryonium located close to the threshold.
There is no experimental support for this result. Apparently it is excluded by the $R_{n/p}$ result in Helium. Also, as shown in
ref\cite{loi20}, it is inconsistent with level shifts in light antiprotonic atoms.
The old pionization experiments used  Helium gas, while the planned PUMA experiment will be performed in vacuum. 
The atomic capture states are likely to be different and possibly the coming experiment will solve the puzzle.

%%%%%%%%%%%%%%%%%%%%%%%%%%%%%%%%
\subsubsection{Atomic level widths  and shifts due to nuclear interactions  }

Finally we give results for atomic level shifts and level widths calculated  with the three discussed potentials. The  method of multiple scattering series summation discussed in reference \cite{loi20} was used. It involves a quasi three-body separation of the system into : (1) = the antiproton (2) = an  interactive nucleon and (3) =  a residual nucleus. For the initial function we use normalized relative (3)-(2) wave functions
\begin{equation}
\label{h1}
 \varphi(r) = [ 1-\exp(-\widetilde{\beta} r ) ]^4~ \frac{ \exp(-\widetilde{\alpha } r)}{r},
 \end{equation}
 where $\widetilde{\alpha}=\sqrt{2\mu_{R,N} E_s}$ and $E_s$ is the separation energy.
 The short distance parameter $\widetilde{\beta}$ is extracted from electron or pion scattering experiments   and fixed to reproduce  the zeros of  charge form-factors.
 For $^3$He, $\widetilde{\alpha }= 0.45 , \widetilde{\beta}=1.753 $~\cite{FEA77}, and for $^4$He$, \widetilde{\alpha }= 0.846, \widetilde{ \beta }=1.20~$\cite{LIM73}, all in fm$^{-1}$ units.

\begin{table}[ht]
\caption{Level shifts $ \epsilon $ and widths  $\Gamma$  in  $2P$ (eV)  and $3D$  (meV) states of  \={p}-He.  Experimental results  come from  Ref. \cite{sch91}.}
\label{LevHe}
\bigskip
\begin{tabular}{|l|l|l|l|l|}\hline
                 & Paris   &   DR       & KW          &   Data  \\\hline
 $^3$He          &          &                &                &        \\
 $\epsilon$(2P)  &1.0      &  7.7           &  -2.1          &  17(4)      \\
 $\Gamma(2P)  $  & 26.2     &  17.5          &  13.0          & 25(9)    \\
 $\Gamma(3D)  $  & 1.59      &  1.48          & 1.57& 2.14(18)     \\\hline
$^4$He           &          &                &                &        \\
 $\epsilon$(2P)  & -1.5       & 10.8          &  1.9           & 18(2)      \\
 $\Gamma(2P)  $  & 26.1      & 18.8          &   26.O         & 45(5)     \\
 $\Gamma(3D)  $  & 1.80      &  1.88         &  1.97          & 2.36(10)     \\\hline 
 \end{tabular}
\end{table}

The level shifts are dominated by the first collision.  
Summation of multiple collisions presented in Appendix \ref{mscat} leads  to results compared  to experimental data.
Results are displayed in Tables \ref{LevHe}  (for He) and \ref{LevH} (for $^2$H). 
As one can see,  all models produce fairly proper level widths,  but they fail badly to reproduce the shifts,  in some cases (Paris or KW)  by one order of magnitude.
 Summation of higher numbers of collisions produces corrections:  small  in $^3$He (Table \ref{LevHe})  and negligible in $^2$H   (Table  \ref{LevH}).   
 On the other hand,  the multiple scattering collisions  in $^4$He strongly  damp the $2P$ level widths. In this atom the  antiproton reaches  much denser parts of the nucleus and  this damping is due to repulsion exerted by the nucleus.     
 An apparent paradox is created in  $^4$He and also in heavier atoms,  \cite{FRI15}. The level shifts are repulsive but fitting  
all data requires an attractive optical potential.   Some possibilities to understand it exist: one  is related to strong  medium effects changing the basic $N\bar{N}$ interactions,  another possibility   is a fairly narrow  $N\bar{N}$ state at about 30-40 MeV  below the threshold. At the moment there is no explanation. 
The upper $3D$ level widths are systematically above the calculated ones. 
The difference is not related to $D$ wave   $\bar{p}N$ contribution which generates  a 0.02 meV extra width. It  may indicate slightly longer-ranged  absorptive $\bar{p}N$  potential. However, these widths are not measured directly and depend on difficult to control cascade calculations. New experiments with high resolution detectors would be helpful to clarify the problem.

\begin{table}[ht]
\caption{ Level shifts $ \epsilon $ and widths  $\Gamma$ in  2P  states of \={p}-$^2$H (in meV). Experimental results come from Refs. \cite{aug99,GOT03}. }\label{LevH}
\bigskip\begin{tabular}{|l|l|l|l|l|}\hline
                 & Paris   &  DR       & KW          &    Data  \\\hline

 $\epsilon$(2P)  & 33      &  84          &  - 48          &  243(33)      \\
 $\Gamma(2P)  $  & 517     &  345           &  380          &  489(30)     \\\hline 
 \end{tabular}
\end{table}
%%%%%%%%%%%%%%%%%%%%%%%%%%%%%%%%%%%
%%%%%%%%%%%%%%%%%%%%%%%%%%%%%%%
\section{Concluding remarks}

The ratio of antiproton-neutron to antiproton-proton captures $R_{n/p}$ was calculated with several  $\bar{N}N$ interaction models.   
In medium and heavy antiprotonic atoms, this ratio is dominated by $P$-waves nucleon-antinucleon interactions and is close to unity in all studied models.  It is consistent with all types of nuclear absorption reactions: pionizations, radioactive final state measurements and level widths of atomic states. 
Thus, one may expect the $R_{n/p}$ to be independent of atomic cascade details and states of nuclear captures.  
As a consequence one expects it to describe nuclear captures also in antiprotonic atoms formed in vacuum.  

The results obtained in He are different. The pionization experiments yield $R_{n/p}$ to be about one half and a comparable  result is obtained only by the  Paris potential and in an $S$-wave nucleon - antinucleon state.  
As shown in this work pionization  data in Helium  are  consistent with dominant  captures in $D$  wave atomic states.  
That indicates an alternative: either a change of Paris potential (suppressing $P$-wave interactions) or in experiments with gas targets 
the majority of nuclear absorptions happening in high $nS,nP$ atomic states.   
However, the second option is much less likely.   
It is the PUMA experiment, which will be performed in vacuum, that may give a more definite  answer. 

The three presented  nucleon-antinucleon potentials fare well in comparison to experimental level widths.  
The Dover-Richard  potential is compatible also with the level shifts. The other two: Paris and Kohno-Weise fail badly in this respect.

\section{Acknowledgements}
This work was supported by the IN2P3-COPIN agreement no 22-154 and realized at National Center for Nuclear Research in Warsaw and at IJClab in Orsay.
We thank the  warm hospitality of both centers and the support of the CNRS/IN2P3 French-Polish agreement.

%%%%%%%%%%%%%%%%%%%%%%%%%%%%%%%%%%%%

\appendix 

%%%%%%%%%%%%%%%%%%%%%%%%%%%%%%%%%%%%%%%%%%%%%%%%
\section{Final state interactions} \label{AA}

Antiprotons  are absorbed by the nucleus  from high angular momentum states.
At nuclear distances the  repulsion due to centrifugal barriers   makes  atomic wave functions   falling  down as $ \Psi(r)   \sim r^L \cdot f(r) $  where $r$ is the radial coordinate and $L$ is the angular momentum.  This makes nuclear captures to happen at distant peripheries, typically at less then $5 \% $ of the central nuclear density. Functions  $ f(r) $ describe deformation of atomic wave functions due  to nuclear interactions and for most of peripheral cases  $ f(r) \simeq 1.$

Antiproton annihilation and related emission of mesons is  a complicated  quantum  process,
The basic transition matrix element for the pionization is
\begin{equation}
\label{ma0} f = < \Psi_{\pi} \Psi_{A-1}|~ t_{\overline{p}N\rightarrow M}~|\Psi_{A} \psi_L>
\end{equation}
where $\Psi$ are wave functions of a final $A-1$  and the initial $A$ nuclei,  $\psi_L $ is the atomic wave function of initial antiproton  and $\Psi_{\pi}$ is  a wave function of the final state mesons.

Multi-body calculations are involved and approximations are necessary.  We refer to refs. \cite{wyc96} and \cite{pis23} for details  and present main formulas below.   The first approximation is the  assumption of completeness of the final states. 
It is allowed by  2 GeV energy release  which goes into kinetic energies of final particles.  It allows to turn phase space integral of mesonic functions  into  a classical integral over the  nuclear density and  the atomic  density overlap.  The reaction rate becomes
\begin{equation}
\label{gama}
\Gamma = \int~d \textbf{r}~  \Sigma_n |\varphi_n(r)|^2  ~   |\Psi_{L}(r)|^2 ~   W(r)~ ~
\end{equation}
where the sum over single particle nucleon states generate nuclear density $\rho(r)$,  and pionization rate  $ W(r) $   is an aggregate describing rate of pionization and effects  of final state mesonic interactions
\begin{equation}
\label{pmiss}
 W(r)  = \int d L^{cm} ~| t_{\overline{p}N\rightarrow M}|^2  |\prod _i |\phi^{-} ( \textbf{r}, \textbf{k})|^2.
\end{equation}
where $\prod _i $ denotes product of all mesonic wave functions $\phi^{-} $ .  

The analysis of  data  in light nuclei requires calculations of charge-exchange probability $\lambda$ and absorption probability $\omega$ per single  emitted meson.  All functions  $ \phi^{-}$  involve final state interactions. The mesons are  formed  on the nuclear surfaces and propagate in all directions. Average mesonic momenta are large $k \sim ~ 400$ MeV/c   and allow  for eikonal  description.  Wave functions for each outgoing pion  are thus given by
\begin{equation}\label{wave}
  \overline{\phi^{-}} ( \textbf{r}, \textbf{k}) =\exp [i \textbf{r} \textbf{k} -i S(\textbf{r}, \textbf{k})]
\end{equation}
and  the transmission probability is calculated as 
\begin{equation}
\label{trans}
 T(\textbf{r}, \textbf{k}) =  |\phi^{-} ( \textbf{r}, \textbf{k})|^2  =\exp [- 2~ Im ~ S(\textbf{r}, \textbf{k})~]
\end{equation}
where the exponential damping of mesonic wave function is determined by  a semi-classical  function $S$ which sums up the damping of mesonic wave as it propagates across the mucleus 

\begin{equation}\label{S}
S(\textbf{r}, \textbf{k}) =  \int_0^\infty d s ~ \left[\sqrt{k^2 -U (\textbf{r}- s ~\widehat{\textbf{k}})} ~ -k  \right]
\end{equation}
where  $U$ is the potential describing  absorption, elastic or inelastic scattering. Coordinate  $\textbf{r}$  refers to the nuclear center.   Boundary conditions are set in such a way that for distant captures  $ 
T(\textbf{r}, \textbf{k})\rightarrow 1 $ as $r\rightarrow\infty $. 
For Klein - Gordon  equation the nuclear optical potentials are determined by cross sections  of elementary reactions with  the normalization 
\begin{equation}\label{U}
U(r) = \sigma ~\rho_f(r)~ \cdot k.
\end{equation}
In practice we expand the root in eq. (\ref{S})  and  consider only inelastic, or absorptive cross sections corresponding to  
\begin{equation}\label{Uabs}
U(r)_{abs}  = -i \sigma_{inel} ~\rho_f(r)~ \cdot k.
\end{equation}

In order to obtain  an average   chance for a single meson to be emitted  one has to calculate  its wave function and  average  the transmission rate  over capture region and over  mesonic  momenta  $k$  and directions of flight.  This involves   integration  over  mesonic phase space $ d L^{cm}$ in each multiplicity and sum over specific transition matrices frequencies  $ | t_{\overline{p}N\rightarrow M}|^2 = M_{\overline{p}N\rightarrow M}$   where frequencies $M$ are  given  in Tables \ref{MP} and \ref{MPT} of  the main text.

The extent of summation depends on the nature of experiment analyzed.
Thus, experiments detecting  cold residual nuclei \cite{lub98}  require  summation of all final mesonic channels and all modes of pionic reactions \cite{wyc96}. The  pionization experiments \cite{bug73,obe20} require all channels but separate:  charge-exchange and  meson absorption modes \cite{pis23}. 
On the other hand, the pionization  experiments in light nuclei \cite{bal87,ben91} detect  specific mesonic states of given number of charged  mesons. Here we follow the last  scenario presented in more detail in ref. \cite{pis23}.

The analysis of data in light nuclei requires calculations of charge-exchange probability $\lambda$ and absorption probability $\omega$ per single produced  meson.   

The  average  transmission probability is calculated as 
\begin{equation}
\label{gama1}
T_\alpha =  \frac{1}{\Theta}  \int~d \textbf{r}~  |\psi_{L}(r)|^2 ~  \Sigma_n   |\varphi_n(r)|^2 ~\int d L^{cm}| M_{\overline{p}N\rightarrow M}|^2 T_\alpha(\textbf{r}, \textbf{k})
\end{equation}
where $ \alpha$ denotes $\pi^+, \pi^-, \pi^0$  and some specific final state , $\int d L^{cm}$ denotes Monte Carlo integration over mesonic phase space for a given multiplicity  discussed in reference \cite{wyc96} and  normalization overlap is 
 \begin{equation}
\label{overlap}
\Theta  =         \int~d \textbf{r}~     |\psi_{L}(r)|^2   \Sigma_n   |\varphi_n(r)|^2  .
\end{equation}
  Probabilities for  captures  (or charge-exchanges ) become $ 1- T_\alpha $.

In the  next subsections we discuss the two  main  absorption modes which determine spectra of final mesonic charges.

%%%%%%%%%%%%%%%%%%%%%%%%%%%%%%%%%
\subsubsection{Charge-exchange processes}

The pion   charge-exchange reaction
\begin{equation}\label{cexap}
\pi^- ~~ p \rightarrow \pi^0 ~~ n
\end{equation}
was studied   up to  $250$ MeV pion kinetic energies in ref. \cite{bre06}. The cross section follows the shape of $ \Delta(1233)$ which dominates the reaction 
\begin{equation}
\label{sigmaex}
 \sigma_{ex} =  48.1  \frac{ (\Gamma/2 )^2 }{ ( \Gamma/2 )^2 + ( T_{\pi} - T_r )^2 }~~~    mb
\end{equation}
where $T_{\pi}$  is a kinetic energy,   $T_r = 167 $ MeV   and the width at the resonance is $ 110$ MeV.
It was extrapolated to  300 MeV  by SAID \cite{bre06}. As we need to extrapolate it further to about 400 MeV   the experimental  $ \Delta(1233)$ profile was followed.
 On  the distant right shoulder of $ \Delta $ our  extrapolation follows   $ \Gamma = 70 +  T_{\pi}/18 $   MeV.

The energies of the pions emitted  in the  $N\overline{N}$  annihilation depend on the number of emitted pions.  Average values are shown  in Table \ref{kinetic}.  

\begin{table}[ht]
\caption{ Average kinetic energies $T_{kin} $  of a single meson formed in a nucleon-antinucleon anihilations with  $ N(\pi)$  mesons emitted.}\label{kinetic}
\bigskip
\begin{tabular}{| l |l|l|l|l|l|l|l|l|l|}\hline
$ N(\pi)$     &  $2 $   &    $ 3 $   &    4 & 5      &  $6 $  &     $ 7 $   &  8   &  9    \\ \hline
$E_{kin} $    &  793    &    482     &  327 &  233   &  171   &     127     &  93  &  67  \\ \hline
\end{tabular}
\end{table}
Maximal cross sections  happen  close to the resonance peak
$T_r \simeq 160-170 $ MeV   that is in the emission of six  mesons.
In the lowest charge channel  of Helium experiment  (one $\pi ^+ \pi^- $ pair supplemented by two  $ \pi^0$)     $N(\pi) =4 $. 
The kinetic energies are much above the resonance and  the charge-exchange probability  $ \lambda $ is very small.  For higher  charge states ($ N(\pi) =5,7 $) measured   in the Helium capture the overlap of mesonic energy spectrum and  $ \Delta(1233)$ is much larger, $ \lambda $  also becomes  larger.

Reaction (\ref{cexap})   and its inverse  are the same by  isospin symmetry  and  differ only  slightly,  by the final phase space. Hence, equation (\ref{sigmaex}) describes  also  other mesonic charge-exchange reactions involved in the final states.

\subsubsection{True two-nucleon absorptions.}\label{true}

We use  theoretically  motivated, energy dependent two-nucleon
capture  approach.  The general understanding is that the $U_{NN}$ term describes two nucleon capture
initiated by the excitations of $\Delta(1233)$.
A direct  model of such  potential was offered by  Muttaz Nuseirat \emph{et al} \cite{mut98}.
This potential is local
\begin{equation}
\label{nn1}
Im ~U_{NN}     = -  \lambda_o
\frac{(\Gamma/2)^2 }{ (E_o- E_\pi)^2 +( \Gamma /2)^2} ~ \rho(r)^2
\end{equation}
with parameters :  $\lambda_o = 46.35   fm ^4 $,  $E_o= 215 $ MeV  and  $\Gamma = 77$ MeV. The  parameters were tested in light nuclei and for
pion kinetic energies $T_{\pi} < 180$  MeV/c, in   $Z=N$ situations.
We find effects of absorptions to be very small in Helium 3 and slightly larger in He 4 with small effect on the final  spectra.

%%%%%%%%%%%%%%%%%%%%%%%%%%%%%%%%%%%%
\section{Fit procedure.}\label{Fit}

Given the data set corresponding to the charge-exchange  transitions $K$=1$\to K'$=0 (i=1),  
$K$=2$\to K'$=1 (i=2) and $K$=3$\to K'$=2 (i=3),
displayed in Tables IV (for $^3$He) and IX (for $^4$He), we aim to determine three real positive numbers $0<w_L<1$  (L=0,1,2)
minimizing the  $\chi^2$ with respect to the experimental data  defined by.

\begin{equation}\label{Chi2_2}
\chi^2(w_0,w_1,w_2) 
={1\over N_d}\sum_{i=1}^3 \frac{ \left[  w_0 \lambda_0^{(i)} + w_1 \lambda_1^{(i)}  + w_2 \lambda_2^{(i)} - \lambda_{Exp}^{(i)} \right]^2 }{ (\Delta \lambda_{Exp}^i)^2 }
\end{equation}
with the constraint
\begin{equation}\label{Constraint}
w_0+w_1+w_2 =1 
\end{equation}
and so the number of  (independent) degrees of freedom $N_d=2$.

\begin{figure}
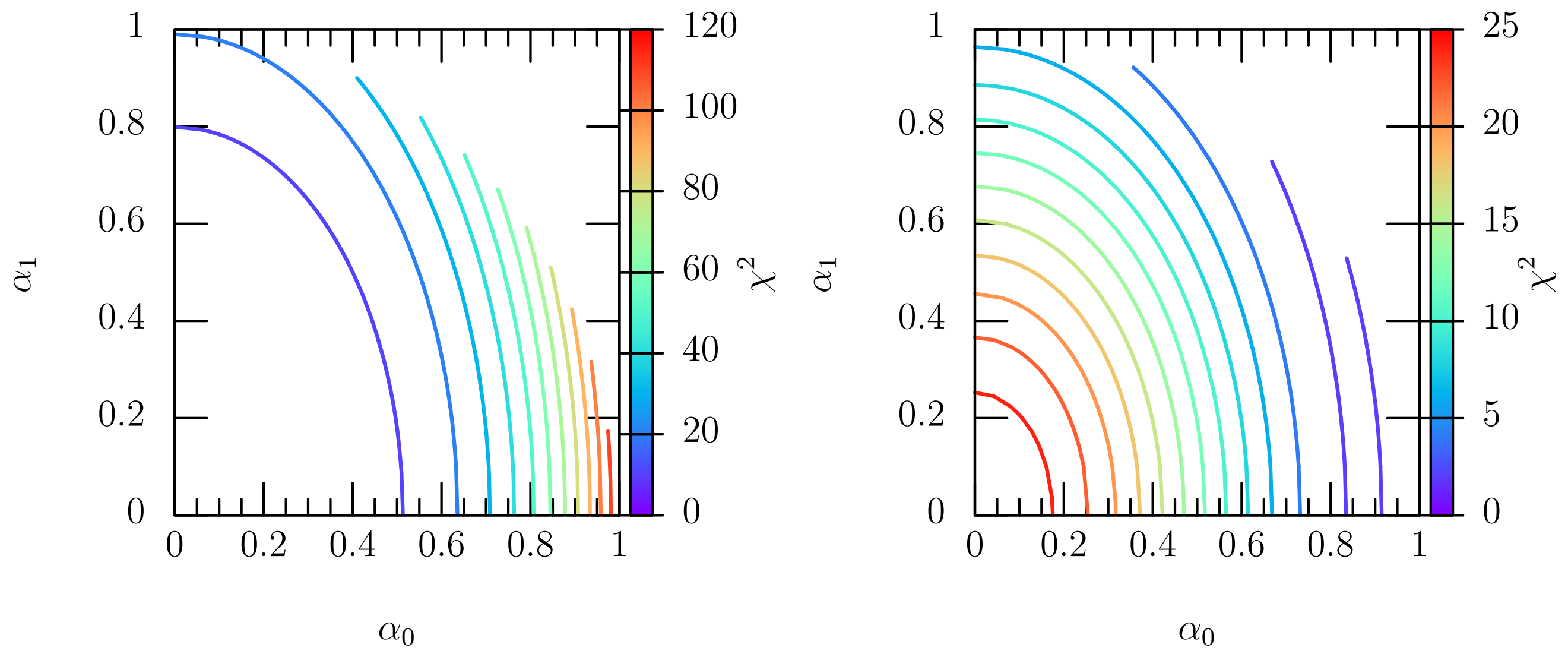

  \centering
  \input{figure-3He.tex}
  \input{figure-4He.tex}
  \caption{Isoline map of $\chi^2$   given by (\ref{Chi2_2}) as a function of $\alpha_L\equiv \sqrt{w_L}$, for $^3$He (left figure) and $^4$He (right figure)}\label{Contour_Chi2}
\end{figure}

Their values correspond to the weight of the L=0,1,2 contribution to the considered transitions, assuming
that these angular momenta exhaust all possibilities.
They have been obtained by numerically searching the minima of (\ref{Chi2_2}) on a two-dimensional plane (positive quadrant) $w_0+w_1=1$
with $w_2$ given by (\ref{Constraint}). Results  are:
\begin{itemize}
\item For $^3$He:
\begin{equation}
 \bar{w}_0=0.00  \qquad \bar{w}_1=0.15 \qquad    \bar{w}_2=0.85  
\end{equation}
with  $\chi^2({\rm minima})=2.33$

\item For $^4$He: 
\begin{equation}
  \bar{w}_0=0.77 \qquad   \bar{w}_1=0.00 \qquad   \bar{w}_2=0.23   
 \end{equation}
 with $\chi^2({\rm minima})=0.90$

\end{itemize}
We have plotted in Fig. \ref{Contour_Chi2} the isoline map of the $\chi^2(w_0,w_1)$ surface
as a function of $\alpha_L\equiv \sqrt{w_L}$  (to obtain quasi-circular isolines).
As one can see the solution is unique (single minima) but the surface is quite flat 
resulting into large uncertainties.

One can estimate the error $\Delta_i$ in each independent parameter $w_i$ with the $\Delta\chi^2$=1 rule, that is
by looking for independent variations of one unit around the minima $(\bar{w}_0,\bar{w}_1,\bar{w}_2)$ in  the $\chi^2$ given by (\ref{Chi2_2})
\[ \chi^2(\bar{w}_0+ \Delta_0,\bar{w}_1+\Delta_1,\bar{w}_2+\Delta_2) = \chi^2(\bar{w}_0,\bar{w}_1,\bar{w}_2)+1\]

The final result of this analysis gives:
\begin{itemize}
 \item     For $^3$He
\[ \bar{w}_0=0.00+ 0.08 \qquad \bar{w}_1=0.15 \pm 0.14 \qquad \bar{w}_2=0.85+0.15 -0.26   \]
\item For $^4$He
\[  \bar{w}_0=0.77 \pm 0.11 \qquad \bar{w}_1=0.00+0.14 \qquad \bar{w}_2=0.23+0.21-0.23  \]
\end{itemize}

%%%%%%%%%%%%%%%%%%%%%%%%%%%%%%%%%%%%%%%%%%%%%%%%%%%%%%%%%%
\section{Summation of multiple-scattering series}\label{mscat}

\noindent
The perturbation expansion for the level shift  is   equivalent to multiple scattering  expansion for meson-nucleus scattering amplitudes.  For high angular momenta discussed here  Coulomb corrections  to the shifts are small and the method of summation used here  follows references \cite{loi20} ,\cite{GRE05}.  In comparison to optical potentials it   prohibits  double scattering on the same nucleon, permits  excitations of the nucleus and allows for correlations.
The solution is built as a multiple-scattering expansion with  nucleon-antinucleon scattering amplitude 
\begin{equation}
\label{fm1}
    t  = ~[ a( E_{cm})  + 3 \overleftarrow{\bigtriangledown } b( E_{cm}) ]  \overrightarrow{\bigtriangledown} 
\end{equation}
where  $a$ is a spin average of $S$-wave  antiproton nucleon  amplitudes  at proper center  of mass energy $E_{cm} $ and  $b$ is  a similar average for $P$ wave amplitudes. 

The scattering series is summed into a quasi geometric series that converges rapidly.
We denote the single scattering term by $<V>$
\begin{equation}
\label{fm2}
   \langle V \rangle = \frac {2\pi}{ \mu}~ \Sigma_i  \langle \Phi \psi_i |  t |\psi \Phi\rangle =\frac {2\pi}{ \mu}~ \Sigma_i ~\langle \Phi ~\psi_i  ~[  a( E_{cm})  + \overleftarrow{\bigtriangledown } b( E_{cm}) ]  \overrightarrow{\bigtriangledown} ~\varphi_i~\Phi \rangle
\end{equation}
where $\Phi$ is an atomic wave function of  antiproton , $\Phi$ is a nuclear (single particle) wave function and $  \varphi $  are wave functions of struck nucleons.
\begin{equation}
\label{fm3}
   \langle VGV\rangle =[\frac {2\pi}{ \mu}]^2~  \Sigma_i  ~ \Sigma_j ~
 \langle \Phi~\varphi_i   [ a + \overleftarrow{\bigtriangledown} b   \overrightarrow{\bigtriangledown}]  ~ G~ [ a+  \overleftarrow{\bigtriangledown } b \overrightarrow{\bigtriangledown}] |\varphi_j~\Phi \rangle.
\end{equation}
The first sum for atomic level shifts   becomes
\begin{equation}
\label{fm4}
\Delta E'   = \frac{\langle V\rangle}{ 1 - \frac{ \langle VGV\rangle}{\langle V\rangle} }.
\end{equation}

Expanding the denominator of equation \ref{fm4} into geometric series yields the first two terms of the  multiple scattering.
To obtain  higher orders, one calculates  triple antiproton -nucleon collisions  and  terms related to struck nucleon - residual nucleus collisions. The latter  may be included into an effective operator $G'$  ( see reference \cite{GRE05} ) and corrects the denominator
\begin{equation}
\label{fm5}
\Delta E''   = \frac{\langle V\rangle}{ 1 + D_1 +  D_2   }
\end{equation}
where  the expansion coefficients are $ D_1 = -\langle VG'V\rangle /     \langle V \rangle $       and    $D_2 =   - (\langle VG'V\rangle  / \langle V  \rangle  ) ^2  \langle VG'VG V\rangle /\langle V\rangle.  $
 In the upper $2P$  levels of interest one finds $|D_1| \sim 0.15; ~|D_2| \sim 0.01 $ and only the first term  $ D_1$ is needed. For the $S$  levels triple collisions have to be calculated to reach a similar precision.
 For detailed formulas and kinematical factors we refer to calculations performed in the  lightest atoms \cite{loi20}.
   The double scattering  term involves a  propagator of the whole system   $G$. It includes  excitation of the system to continuum. In principle,  refined effects like correlations and nuclear excitations  are hidden there. However, since   $  \langle VGV\rangle /   \langle V\rangle $ is small in atomic P,D states we stop   at expression $ \Delta E'$.

%%%%%%%%%%%%%%%%%%%%%%%%%%%%%%%%%%%%%%%%%%%%%%%%%%%%%%%%%%%%%

%\bibitem{alq02} M.Alquadi and W.R.  Gibbs   "Final state interaction in pion production from nuclei " ,Phys. Rev. C 66(2002))064604

%%%%%%%%%%%%%%%%%%%%%%%

\end{document}